\documentclass[prd,amsmath,amssymb,superscriptaddress,floatfix,nofootinbib,10pt]{revtex4}
\usepackage{times}
\usepackage{amssymb,amsbsy,amsmath,amsfonts}
\usepackage{graphicx}
\usepackage{float}
\usepackage{color}
\usepackage{morefloats}
\usepackage{rotating}
\usepackage{srcltx}
\usepackage{slashed}
\usepackage{multirow}
\usepackage{verbatim}
\usepackage{hyperref}
\usepackage{tabularx}
\usepackage{bm}
\allowdisplaybreaks[4]

\usepackage{bbding}
\usepackage{threeparttable}

\usepackage{mathrsfs}
\DeclareUnicodeCharacter{2212}{\ensuremath{-}}

\newcommand{\PreserveBackslash}[1]{\let\temp=\\#1\let\\=\temp}
\newcolumntype{C}[1]{>{\PreserveBackslash\centering}p{#1}}
\newcolumntype{R}[1]{>{\PreserveBackslash\raggedleft}p{#1}}
\newcolumntype{L}[1]{>{\PreserveBackslash\raggedright}p{#1}}

\begin{document}

\title{ $P_c(4440)$ and $P_c(4457)$ decay into $\bar{D}\Sigma_c$ and $\bar{D}\Lambda_c$ and the spin of the $P_c$ states}

\author{Zi-Ying Yang}
\affiliation{School of Physics, Beihang University, Beijing, 102206, China}
\affiliation{Departamento de Física Teórica and IFIC, Centro Mixto Universidad de Valencia-CSIC Institutos de Investigación de Paterna, 46071 Valencia, Spain}

\author{Jing Song}
\email[]{Song-Jing@buaa.edu.cn}
\affiliation{School of Physics, Beihang University, Beijing, 102206, China}
\affiliation{Departamento de Física Teórica and IFIC, Centro Mixto Universidad de Valencia-CSIC Institutos de Investigación de Paterna, 46071 Valencia, Spain}

\author{Wei-Hong Liang}
\email[]{liangwh@gxnu.edu.cn}
\affiliation{Department of Physics, Guangxi Normal University, Guilin 541004, China}
\affiliation{Guangxi Key Laboratory of Nuclear Physics and Technology,
Guangxi Normal University, Guilin 541004, China}

\author{ Eulogio Oset}
\email[]{oset@ific.uv.es}
\affiliation{Departamento de Física Teórica and IFIC, Centro Mixto Universidad de Valencia-CSIC Institutos de Investigación de Paterna, 46071 Valencia, Spain}
\affiliation{Department of Physics, Guangxi Normal University, Guilin 541004, China}

\begin{abstract}
   We address the issue of the width and spin assignment of the $P_c(4440)$ and $P_c(4457)$ pentaquark states. We calculate the partial decays widths of the particles into $J/\psi N$,  $\bar{D} \Sigma_c$ and $\bar{D} \Lambda_c$ with the first one obtained within a unitary approach with the coupled channels of $J/\psi N$ and $\bar{D}^* \Sigma_c$ with the interaction driven by vector meson exchange, and the last two by means of triangle diagrams involving pion exchange which break the spin degeneracy of the vector meson exchange. The widths obtained   depend much on the spin of the particles and we study the possible scenarios for spin assignment, favoring the $1/2^-$ , $3/2^-$ order. We predict values for decays in different channels and show that their determination would allow to decide unambiguously  the spin of the states. 

\end{abstract}

%\pacs{13.75.Ev,12.39.Fe,21.30.Fe}
%\keywords{}

\maketitle

\section{introduction}
The discovery of the $P_c$ states in \cite{LHCb:2015yax}, updated in \cite{LHCb:2019kea} by the LHCb Collaboration, has triggered an enormous amount of work. It is interesting to mention that starting from the work of \cite{Wu:2010jy,Wu:2010vk} many works had done predictions for such states \cite{Wang:2011rga,Yang:2011wz,Wu:2012md,Xiao:2013yca,Li:2014gra,Chen:2015loa,Karliner:2015ina}.
Numerous theoretical interpretations on the
nature of the pentaquarks followed these discoveries, including hadronic molecules \cite{Zhang:2023czx,Liu:2019tjn,Chen:2016qju,Du:2021fmf,Du:2019pij, 
Chen:2019bip,Chen:2019asm,Guo:2019twa,He:2019ify,Guo:2019kdc,Shimizu:2019ptd,Xiao:2019pjg,Xiao:2019aya,Wang:2019nwt,Meng:2019ilv,Wu:2019adv,Xiao:2019gjd,Voloshin:2019aut,Sakai:2019qph,Wang:2019hyc,Yamaguchi:2019seo,Liu:2019zvb,Lin:2019qiv,Wang:2019ato,Gutsche:2019mkg,Burns:2019iih,Wang:2019spc,Xu:2020gjl,Kuang:2020bnk,Peng:2020xrf,Peng:2020gwk,Xiao:2020frg,Dong:2021juy,Peng:2021hkr},
compact pentaquark states \cite{Kuang:2020bnk, Ali:2019npk,Zhu:2019iwm,Wang:2019got,Giron:2019bcs,Cheng:2019obk,Stancu:2019qga}, hadro-charmonia \cite{Eides:2015dtr,Eides:2019tgv,Ferretti:2018ojb}, and cusp effects \cite{Kuang:2020bnk}.
Particular interest have attracted the $P_c(4440)$ and $P_c(4457)$ states, which in most of the molecular picture works appear as degenerate.
In \cite{Liu:2019tjn,Guo:2019kdc}   the $P_c(4440)$ and $P_c(4457)$ are associated to a molecular state of $\bar{D}^*\Sigma_c$ with $1/2^-$, $3/2^-$ but the particular spin of each state is ambiguous. Works based on Heavy Quark Spin Symmetry {alone \cite{Chen:2019asm, He:2019ify, Liu:2019zvb, Du:2019pij, Du:2021bgb} cannot} decide which is the spin of each particle. 
 In \cite{Du:2019pij,Du:2021fmf} three schemes are proposed, in some of which there is ambiguity about the spin assignment, but a preference is shown for $3/2^-$, $1/2^-$ for the $P_c(4440)$ and $P_c(4457)$ states respectively when pion exchange is considered and requirements about renormalizability are imposed. Most works consider the pion exchange potential as static (three momentum dependent)  (not in \cite{Du:2019pij,Du:2021fmf}) to be included in the Lippmann Schwinger equation. Our formalism considers dynamical pions, where the full propagator of the pion is considered in loops evaluated with a fourth fold integration. 
 In \cite{Liu:2019zvb} using the one-boson-exchange model the authors predict the preferred
quantum numbers of the $P_c(4440)$ and $P_c(4457)$ molecular pentaquarks to be $3/2^-$ and $1/2^-$, respectively, but in \cite{Pan:2019skd} two scenarios are shown in which either assignment of the spins is preferred. In \cite{Chen:2019asm,Liu:2019zvb,He:2019ify,Xiao:2019aya,Meng:2019ilv} the $1/2^-$, $3/2^-$ assignment is preferred, while in \cite{Yamaguchi:2019seo,PavonValderrama:2019nbk,Liu:2019zvb} the opposite order is preferred. Different reactions are suggested in \cite{Wang:2019spc} to discriminate the spin of the particles. 
In \cite{Xu:2020gjl} the $1/2^-$ assignment to the $P_c(4440)$ is favored, while the $P_c(4457)$ would involve a mixture of spins. Ambiguities in the spin of the states are also discussed in \cite{Peng:2020gwk}. The $1/2^-$, $3/2^-$ assignment is also preferred in \cite{Xiao:2020frg}. The $3/2^-$, $1/2^-$ order is also preferred in \cite{Peng:2020hql}, while the opposite order is lightly favored in \cite{Liu:2020hcv}. 
In \cite{Yalikun:2021bfm} a discussion is done on how the results depend on the way the pion exchange is treated, if the implicit delta function is or not considered. On the other hand, demanding consistency of the $P_c$ states with the $T_{cc}$ state, {in Ref. \cite{Chen:2021cfl} } the preferred assignment is $1/2^-$, $3/2^-$. 
%So is the conclusion of \R{ Ref. \cite{ollerzhihui}} where it is mentioned that the inclusion of the $\Xi_c^{'}\bar{D}$ channel has some effect on the splitting of the states. 
 Lattice QCD results \cite{Xing:2022ijm} succeed in getting bound states of $\bar{D}^*\Sigma_c$ molecular nature, but are unable to distinguish the spin.  By contrast, the $3/2^-$, $1/2^-$ assignment is preferred in \cite{Yang:2022ezl}. 
In \cite{Zhang:2023czx}, using machine learning techniques, but a pion-less approach, the authors favor  $1/2^-$, $3/2^-$.  The same assignment is also favored in Ref. \cite{Lin:2023ihj}. 
In Ref. \cite{Liu:2023wfo} the authors propose to measure correlation functions  of the meson-baryon components in order to { elucidate } the spin of the particles.
Similarly, the use of the effective range expansion has been advocated in~\cite{Peng:2024yzn}
 to  discriminate the spin of the states.
 The measurement of three body decay channels has also been advocated to determine the spin of the states in \cite{Liu:2024ugt}. 

%In the $\Sigma_c^{(*)} \bar{D}^{(*)}$ hadronic molecular picture, the $P_c(4440)$ and $P_c(4457)$ are primarily associated with the $\Sigma_c \bar{D}^*$ system, and their spin-parity assignments remain uncertain. Two potential solutions have been proposed: either $J^P = 1/2^-$, $3/2^-$ (solution A) or $J^P = 3/2^-$, $1/2^-$ (solution B)~\cite{Liu:2019tjn, Du:2021fmf, Du:2019pij}. The degenerate nature of these states, along with their narrow widths, provides further support for the molecular interpretation~\cite{Liu:2019tjn, Du:2021fmf}. 

Alternative interpretations suggest that these states may be compact pentaquarks. In the diquark-diquark-antiquark model, the $P_c(4440)$ and $P_c(4457)$ are viewed as tightly bound $uudc\bar{c}$ systems, with possible spin-parity values of $J^P = 1/2^-$ or $3/2^-$~\cite{Shi:2021wyt, Wang:2020eep, Azizi:2021utt}. This compact pentaquark interpretation is supported by QCD sum rule and effective Lagrangian calculations, which also emphasize the importance of decay channels like $P_{cs}(4459) \to J/\psi \Lambda$~\cite{Shi:2021wyt, Wang:2020eep, Azizi:2021utt}. 

Despite these developments, the exact nature of the $P_c(4440)$ and $P_c(4457)$ remains elusive. Effective field theory and phenomenological studies show that the mass spectrum alone cannot conclusively determine the spin-parity quantum numbers of these states~\cite{Chen:2019asm,Liu:2023wfo}. As a result, further experimental observations, including measurements of decay widths, angular distributions, and production mechanisms, are needed to definitively resolve their spin-parity assignments and clarify their underlying structure~\cite{PavonValderrama:2019nbk,Chen:2019asm,Zhang:2023czx,Liu:2023wfo}.

    As we can see, the issue of the spin of the  $P_c(4440)$ and $P_c(4457)$ has attracted and continues attracting much attention from the theory side. And the claims of the different works are more or less split into one spin assignment and the opposite.  In this situation we present here a different approach in which a novel decomposition of the   states is made and we determine the decay width of the states into $\bar{D}\Sigma_c$ and $\bar{D}\Lambda_c$, which is found larger for the  $1/2^-$ state. The evaluation of these widths is done using triangle loops involving pion  exchange, which are calculated by means of a four dimensional loop integration considering the full pion propagator.  Then we perform a coupled channel calculation of the $\bar{D}^*\Sigma_c$ and $J/\psi N$ channels with the interaction given by an extension of the local hidden gauge approach, which were found dominant in former studies, and include then the selfenergy of the states due to their coupling to the $\bar{D}\Sigma_c$ and $\bar{D}\Lambda_c$ components. We find that the $1/2^-$ state is more bound and has a larger width, while the $3/2^-$ state is less bound and has a smaller width. 
    Yet, a quantitative description  of the mass difference  between the states cannot be accomplished within the scheme.
We evaluate the different decay widths in two scenarios with different  assignments of $1/2^-,~3/2^-$ for the  $P_c(4440)$, $P_c(4457)$   and favor the order $1/2^-,~3/2^-$. We show that the measurement of the  $\bar{D}\Sigma_c$, $\bar{D}\Lambda_c$   decays would bring light into the problem, allowing a clear assignment.

%    With the help of just one parameter to regularize the loops, we are able to find a satisfactory description of the two $P_c(4440)$ and $P_c(4457)$ states, mass and width, in which the association of the spin is  $1/2^-$, $3/2^-$ for the $P_c(4440)$ and $P_c(4457)$ respectively, comes naturally. 
    
% In this work, we revisit the spin-parity assignments of the $P_c(4440)$ and $P_c(4457)$ states, considering both the hadronic molecular and compact pentaquark models. We aim to use recent experimental data and theoretical advancements to constrain the underlying structure of these intriguing states and resolve the ambiguity surrounding their quantum numbers. Our analysis incorporates the interplay of the $\bar{D}\Sigma_c$ and $\bar{D}\Lambda_c$ systems to explore their spin-parity configurations, providing further insight into the nature of these pentaquark candidates.

%\newpage
\section{Formalism}
To calculate the widths for the $P_c$ decay into $\bar{D}\Sigma_c$ and $\bar{D}\Lambda_c$, we  must consider the isospin combinations of the states. For this we need to express our phase convention 
$\begin{Bmatrix} {D}^{*+} \\ -D^{*0} \end{Bmatrix}, \begin{Bmatrix} \bar{D}^{*0} \\ D^{*-}  \end{Bmatrix}, \begin{Bmatrix} \Sigma_c^{++} \\ \Sigma_c^{+} \\\Sigma_c^{0}    \end{Bmatrix}$.
%With the $I=1/2$ of  $\Sigma_c\bar{D}^*$ to the Clobsh-Gorden coefficients, and 
Then the isospin wave function for the $\Sigma_c\bar{D}^*$ state is given by

\begin{align}\label{1}
\vert \Sigma_c\bar{D}^*,~ I=1/2,~ I_3=1/2 \rangle &=  \sqrt{\frac{2}{3}} ~\Sigma_c^{++} {D}^{*-}   - \sqrt{\frac{1}{3}}~ \Sigma_c^{+} {\bar{D}}^{*0}.
\end{align}

\begin{figure}[H]
  \centering
   \includegraphics[width=0.42\textwidth]{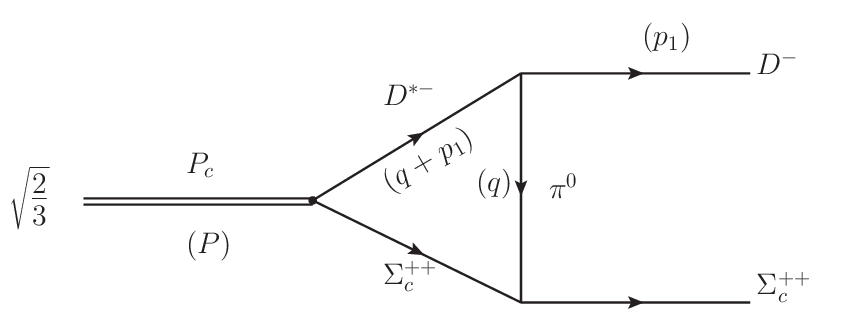}
   \includegraphics[width=0.46\textwidth]{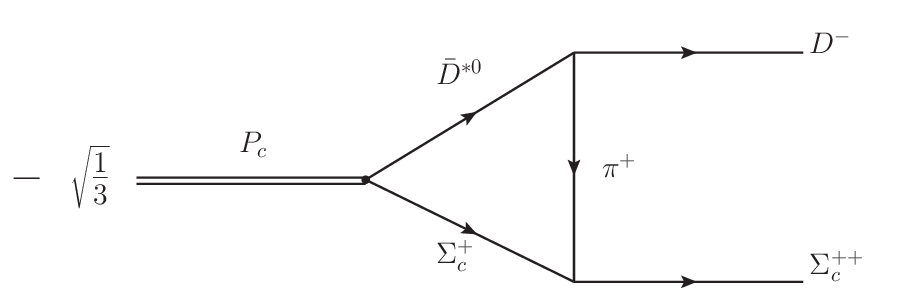}
   \includegraphics[width=0.42\textwidth]{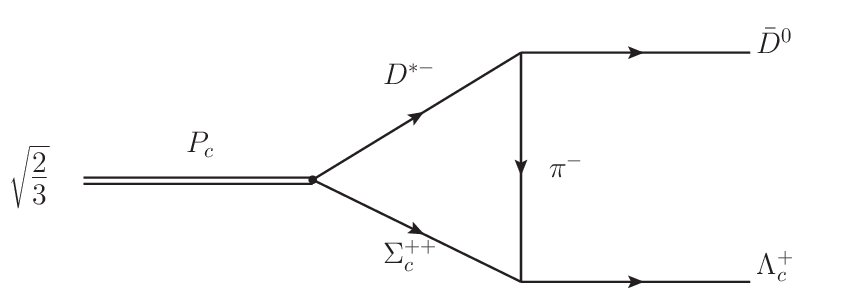}
   \includegraphics[width=0.46\textwidth]{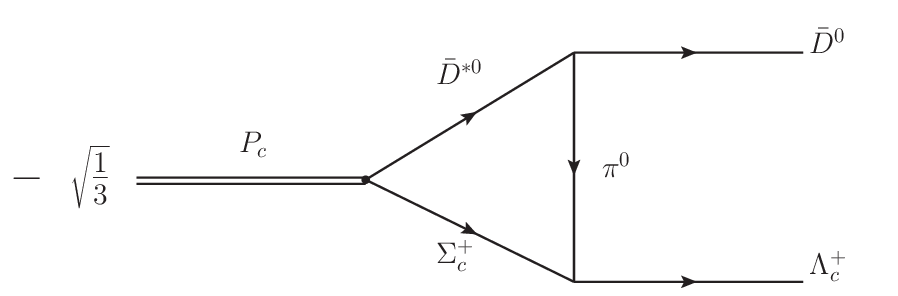}   
  \caption{Mechanism of the reaction \( P_c \to \Sigma_c D ~~(\Lambda_c D)\). In parenthesis the momenta of the particles.}
   \label{feynDiag2}
\end{figure}

The decay widths for $P_c\to{D}^{-}\Sigma_c^{++} $ and $P_c\to\bar{D}^{0}\Lambda_c^{+} $  are evaluated using the triangle diagrams of Fig.~\ref{feynDiag2}.
The evaluation of the diagrams of Fig.~\ref{feynDiag2} requires the knowledge of the couplings of the $P_c$ to the $\bar{D}^*\Sigma_c$ components, the vector-pseudoscalar-pseudoscalar ($VPP$) vertices and the pion-baryon-baryon ($\pi BB$) vertex. We address them one by one. In the first  place we look at how the couplings of the $P_c$ to the $\bar{D}^*\Sigma_c$ components are calculated.
\subsection { Determination of the $P_c \to \bar{D}^*\Sigma_c$ coupling}
\par In order to calculate the $P_c \to \bar{D}^*\Sigma_c$ coupling we rely upon the work of Refs.~\cite{Xiao:2013yca,Xiao:2019aya} where heavy quark symmetry arguments are combined with elements of an extension of the local hidden gauge approach~\cite{Bando:1984ej,bando1988nonlinear,Harada:2003jx,Meissner:1987ge,Nagahiro:2008cv} exchanging vector mesons, to obtain the $\bar{D}^*\Sigma_c$ interaction with its coupled channels. We observe that for the states that we are concerned now, the main coupling of the resonance is to  $\bar{D}^*\Sigma_c$, by large, and then the only other relevant channel is $J/ \psi N$, the channel where the state is observed.

\par
\begin{figure}[h]
    \centering
    \includegraphics[width=0.4\textwidth]{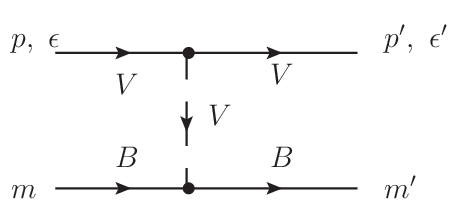}
    \caption{Diagrammatic representation of the $VB\to VB$ interaction, through the exchange of vector mesons. $p,~p'$ and $\epsilon,~\epsilon'$ stand for the momenta and polarizations of the external vectors and $m,~m'$ for the spin third components.}
    \label{fig3}
\end{figure}

 Then we reconstruct the generation of the $P_c$ state with only these two channels within this approach, and we get the $1/2^-,3/2^-$ states degenerate. The mechanism to generate the interaction is depicted in {Fig.~\ref{fig3}.} The upper vertex, $VVV$, is constructed using the Lagrangian

\begin{equation}\label{LGRvvv}
\mathcal{L}_{VVV} = i g \left\langle \left(  V^{\mu} \partial_{\nu} V_{\mu} - \partial_{\nu} V^{\mu} V_{\mu} \right) V^{\nu} \right\rangle,
\end{equation}
with $g=\frac{M_{V}}{2f} ~ (M_{V}=800~\text{MeV}, f=93~\text{MeV})$, and the lower vertex $VBB$ is evaluated using the quark wave functions for mesons and baryons~\cite{Capstick:1986ter,Roberts:2007ni,Wang:2022aga}. We obtain the spin independent transition potential
\begin{equation}\label{a}
    V_{ij}=-\frac{1}{4f^2}C_{ij}(p^0+p^{'0})\Vec{\epsilon}~\Vec{\epsilon}~',
\end{equation}
where $p^0,~p^{'0}(\Vec{\epsilon},~\Vec{\epsilon}~')$ are the energies  ( polarization vectors ) of the initial and final mesons respectively, and $C_{ij}$ is a matrix with the channels $J/ \psi N~(1)$ and $\bar{D}^*\Sigma_c(I=1/2) ~(2)$, given by \footnote{The diagonal terms in $C_{ij}$ are the same as in Refs.~\cite{Xiao:2013yca,Xiao:2019aya} but the non-diagonal term is different. Here we obtain it by explicitly exchanging a $D^*$ meson, while in Refs.~\cite{Xiao:2013yca,Xiao:2019aya} one relies upon heavy quark spin symmetry, which holds for the terms that exchange a light vector, where the heavy quarks are spectators, but nor for those where a heavy vector is exchanged.}
\begin{equation}
C_{ij} =\left(\begin{array}{cc}
0 & -\frac{1}{\sqrt{2}}\frac{m_{\rho}^{2}}{m_{D^*}^{2}} \\
-\frac{1}{\sqrt{2}}\frac{m_{\rho}^{2}}{m_{D^*}^{2}} & 1
\end{array}\right).
\end{equation}
%%%%%%%%%%%%%%%%%%%%%%%%%%%%%%%%%%%%%%%%%%%%%%
The scattering matrix is then given by
\begin{equation}\label{c}
    T=[1-VG]^{-1}V,
\end{equation}
with $G$ the meson-baryon loop function
\begin{equation}\label{Gfun}
    G(s)=\int_{|\Vec{q}~|<q_\mathrm{max}}2M_B\frac{d^3q}{(2\pi)^3} \frac{\omega_1(q)+\omega_2(q)}{2\omega_1(q)\omega_2(q)} \frac{1}{s-\left[\omega_1(q)+\omega_2(q)\right]^2+i\varepsilon},
\end{equation}
where $\omega_j=\sqrt{m_j^2+\Vec{q}^{~2}}$, $j$ for the vector and the baryon, and $M_B$ the baryon mass.
\par The value of $q_\mathrm{max}$ is a regulator that reflects the range of the interaction~\cite{Song:2022yvz}. The couplings are obtained from the $T_{ij}$ matrix at  the pole where it behaves as
\begin{equation}
    T_{ij}\simeq \frac{g_{i}g_{j}}{\sqrt{s}-M_R},
\end{equation}
%%%%%%%%%%%%%%%%%%%%%%%%%%%%%%%%%%%%%%%%%%%%%%%
where $M_R$ is the position of the pole in the complex plane, while $\text{Re}(M_R)$ is the mass of the resonance and $\text{Im}(M_R)$ is half the width.
\par In Table~\ref{POLE_LIANG} we write the values of $M_R$ and the couplings obtained for different values of $q_\mathrm{max}$. We can see that a mass of the state around 4440~\text{MeV} is obtained with $q_\mathrm{max}=580$~MeV, and 4457~\text{MeV} is obtained with $q_\mathrm{max}=450$~MeV. The Table also gives the couplings of the resonance to $\bar{D}^*\Sigma_c$ which will be used in the evaluation of the triangle diagrams.
%In Table~\ref{POLE_LIANG}, we present the masses and widths of the states obtained, along with their couplings to each channel obtained with the $J/\psi N$ and $\bar{D}^*\Sigma_c$  and channels. 
%The calculations are performed using different values of $q_\mathrm{max}$, in particular $q_\mathrm{max}=450$~MeV and $q_\mathrm{max}=580$~~\text{MeV}, which are (decay widths) found to be appropriate for the $P_c(4440)$ and $P_c(4457)$ states. 
The widths of these states, coming from the decay into $J/\psi N$,  are determined as twice the imaginary part of the pole position. 
%Based on the couplings listed in Table~\ref{POLE_LIANG}, the $\bar{D}\Sigma_c$ channel emerges as the most favorable for observation. 
%This indicates that the $\bar{D}\Sigma_c$ channel has a significant contribution to both the $P_c(4440)$ and $P_c(4457)$ states.

\begin{table}[H]
%\footnotesize
\centering
\caption{ Pole position and couplings  $g_i$ \text { of the } $P_c$ state. }
\label{POLE_LIANG}
\setlength{\tabcolsep}{38pt}
\begin{tabular}{cccc}
\hline \hline
$q_{\text {max }}~[\mathrm{MeV}]$ & $\text { Poles }[\mathrm{MeV}]$ & $g_{J / \psi N}$ & $g_{\bar{D}^*\Sigma_c}$ \\
\hline 
620 & $4432.10+i\,2.56$ & $0.436-i\, 0.003$ & $-3.174+i\,0.112$ \\
600 & $4436.50+i\,2.30$ & $0.411-i\, 0.003$ & $-3.002+i\,0.110$ \\
580 & $4440.48+i\,2.05$ & $0.388-i\, 0.003$ & $-2.830+i\, 0.107$ \\
550 & $4445.70+i\,1.70$ & $0.353-i\, 0.004$ & $-2.572+i\, 0.104$ \\
500 & $4452.55+i\, 1.19$ & $0.294-i\, 0.005$ & $-2.144+i\,0.098$ \\
450 & $4457.30+i\, 0.76$ & $0.235-i\, 0.006$ & $-1.714+i\, 0.093$ \\
400 & $4460.25+i\, 0.42$ & $0.174-i\, 0.007$ & $-1.272+i\, 0.092$ \\
330 & $4461.95+i\, 0.07$ & $0.075-i\, 0.016$ & $-0.543+i\, 0.132$ \\
\hline \hline
\end{tabular}
\end{table}

\par As we can see in Table I, the width of the state coming from $P_c \to J/\psi N$, which is around 4~\text{MeV} at 4440~\text{MeV} and around 1.5~\text{MeV} at 4457~\text{MeV}, does not account for the central values of the  whole experimental width as given by~\cite{LHCb:2015yax,LHCb:2019kea} 
\begin{equation}
    \begin{aligned}
        &\Gamma(P_c(4440))=20 \pm 4.9^{+8.7}_{-10.1} ~\text{MeV},\\
        &\Gamma(P_c(4457))=6.4 \pm 2^{+5.7}_{-1.9} ~\text{MeV}.
    \end{aligned}
\end{equation}

\subsection{Pion exchange}
\begin{figure}[h]
    \centering
    \includegraphics[width=0.4\textwidth]{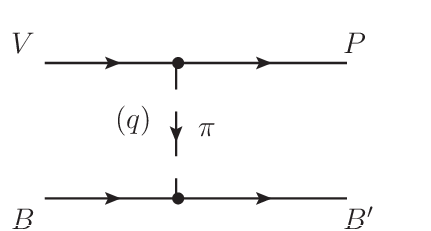}
    \caption{Diagrammatic representation of the interaction $VB\to P B^\prime$ through the exchange of a pion.
                The $V (P)$ and $B (B^\prime)$ are the initial (final) meson and baryon states, respectively.}
    \label{figMB}
\end{figure}

%As mentioned above, we use vector exchange from the extension of the local hidden gauge approach between mesons 
%and baryons as shown in the Fig.~\ref{figMB}. 
Next we turn to the $VB\to PB'$ interaction depicted in Fig.~\ref{figMB}.
The $VPP$ vertex ($V\equiv$ vector, $P\equiv$ pseudoscalar), is described by the following Lagrangian
\begin{align}\label{VPP}
    \mathcal{L}_{\mathrm{VPP}} &= -i g\left\langle\left[P, \partial_{\mu} P\right] V^{\mu}\right\rangle.  
\end{align}
%The coupling  $g=\frac{m_V}{2f_\pi}$ with $m_V=800$~\text{MeV} and the pion decay constant $f_\pi=93$~\text{MeV}. 
The $P$ or $V$ matrices in Eq.~(\ref{VPP})
 are the $q_i\bar{q}_j$ matrices written in terms of mesons and the symbol $\langle\cdot \cdot \cdot\rangle$ means the 
trace of the matrices.

The matrices $P$ and $V$ that we need are given by 
\begin{equation}%%%%%%%
 \label{eq:matP_charm} 
  P = \begin{pmatrix}
          \frac{1}{\sqrt{2}}\pi^0 + \frac{1}{\sqrt{3}} \eta + \frac{1}{\sqrt{6}}\eta' & \pi^+ & K^+ & \bar{D}^0 \\
          \pi^- & -\frac{1}{\sqrt{2}}\pi^0 + \frac{1}{\sqrt{3}} \eta + \frac{1}{\sqrt{6}}\eta' & K^0 & D^- \\
          K^- & \bar{K}^0 & -\frac{1}{\sqrt{3}} \eta + \sqrt{\frac{2}{3}}\eta' & D_s^- \\
          D^0  & D^+ & D_s^+ & \eta_c
       \end{pmatrix},
\end{equation}
\begin{equation}
  \label{eq:matV_charm}
   V = \begin{pmatrix}
             \frac{1}{\sqrt{2}}\rho^0 + \frac{1}{\sqrt{2}} \omega & \rho^+ & K^{* +} & \bar{D}^{* 0} \\
             \rho^- & -\frac{1}{\sqrt{2}}\rho^0 + \frac{1}{\sqrt{2}} \omega  & K^{* 0} & {D}^{* -} \\
             K^{* -} & \bar{K}^{* 0}  & \phi & D_s^{* -} \\
             D^{* 0} & D^{* +} & D_s^{* +} & J/\psi
        \end{pmatrix}.
\end{equation}%%%%%%%
Evaluating the upper vertices of Fig.~\ref{feynDiag2}, we obtain

\begin{align}\label{uppervet}
 &        -it=i g \, \frac{1}{\sqrt{2}}\vec{\epsilon} \cdot ( \vec{q}-\vec{p}_1), ~~\qquad\qquad \text{for~~ $D^{*-} \to\pi^0 D^-$},\\\nonumber
 &        -it=-i g \,\vec{\epsilon} \cdot ( \vec{q}-\vec{p}_1), ~~\quad\qquad\qquad \text{for~~ $\bar{D}^{*0} \to\pi^+ {D}^-$},\\\nonumber 
 &        -it=-i g \, \vec{\epsilon} \cdot ( \vec{q}-\vec{p}_1), ~~\quad\qquad\qquad \text{for~~ ${D}^{*-} \to\pi^- \bar{D}^0$},\\\nonumber 
 &        -it=-i g \, \frac{1}{\sqrt{2}}\vec{\epsilon} \cdot ( \vec{q}-\vec{p}_1), \qquad\qquad \text{for~~ $\bar{D}^{*0} \to\pi^0 \bar{D}^0$},
\end{align}
where $\epsilon^0$  is neglected consistently with Eq.~(\ref{LGRvvv}), which implies small vector meson momenta compared to the vector meson mass.

The lower vertices $\pi\Sigma_c\Sigma_c$, $\pi\Sigma_c\Lambda_c$   were evaluated in Ref.~\cite{Uchino:2015uha} by analogy to $\pi\Sigma\Sigma$ and $\pi\Sigma\Lambda$, since the charmed quarks are {spectators} in the   $\pi\Sigma_c\Sigma_c$, $\pi\Sigma_c\Lambda_c$     vertices and the strange quarks are also {spectators} in the $SU(3)$ sector. Here we use wave functions of $\Sigma_c$ and $\Lambda_c$ where the charmed quark is singled out and the symmetry of the wave function is demanded from the light quark, as done in Refs.~\cite{Capstick:1986ter,Roberts:2007ni}. Thus, 
\begin{align}\label{wavefunc}
   & \Sigma_c^{++} = uuc~\chi_{MS}, \qquad \Sigma_c^+ = \frac{1}{\sqrt{2}}(ud+du)c~\chi_{MS}, \qquad \Sigma_c^{0} = ddc~\chi_{MS},\\\nonumber
   & \Lambda_c^+ = \frac{1}{\sqrt{2}}(ud-du)c~\chi_{MA},
\end{align}
with $\chi_{MS},~\chi_{MA}$ the mixed symmetric, mixed antisymmetric spin wave functions for the three quarks~\cite{Close:1979bt}.
 To evaluate the $\pi$ coupling to these baryons, we   take a coupling of the pions to the quarks as 
\begin{align}
   -it_{\pi qq}=\frac{\tilde{f}}{m_\pi}\sum_i\vec{\sigma_i}\vec{q}\tau_i^\lambda,
\end{align}
where the sum over $i$ is with respect to all quarks and $\tau_i^{\lambda}$ is the isospin $1/2$ Pauli matrix with $\lambda$ the index in the spherical basis, and  $\vec{\sigma}$  the ordinary spin matrix for spin $1/2$. Applying this operator to $\vec{q}$ in the $z$ direction and the $\pi^0 pp$ vertex  ($p\equiv proton $, $\vec{\sigma}\vec{q}\to {\sigma_z}{q}$, $\tau^\lambda\to \tau_z$)  and using 
\begin{align}
    \Psi_p=\frac{1}{\sqrt{2}}~(\phi_{MS}\chi_{MS}+\phi_{MA}\chi_{MA}),
\end{align}
with the flavor $\phi_{MS},~\phi_{MA}$, and spin $\chi_{MS},~\chi_{MA}$ of Ref.~\cite{Close:1979bt},
we find a relationship of $\tilde{f}$ to the $f_{\pi N}$ couplings of $\pi^0 pp$, which at the {macroscopic} level is written as 
\begin{align}
     -it_{\pi^0 pp}=\frac{f_{\pi N}}{m_\pi}\vec{\sigma}\vec{q},
\end{align}
with ${f_{\pi N}}=1.00$ and $m_\pi$ the pion mass.
Then we find that the operator at the quark level to be sandwiched with the baryons is 
\begin{equation}
   \tilde{V}_\pi \equiv \frac{3}{5}\frac{f_{\pi N}}{m_\pi}\sqrt{2} \sum_i \left\{ 
         \begin{aligned}
           & \qquad~~ u\bar{d} \\
           & \frac{1}{\sqrt{2}} (u\bar{u} - d\bar{d}) \\
           & \qquad~~ d\bar{u}\\
         \end{aligned} \right\}_i\sigma_{z,~ i},
     \label{eq:L_VBB}
\end{equation}
where $\{\}$ stands for the quark components of  $\left\{ 
         \begin{aligned}
           & \pi^+ \\
           & \pi^0 \\
           & \pi^-\\
         \end{aligned} \right\}$, and the coupling $\pi BB'$ is then given by 
\begin{align}
     -it_{\pi BB'}=\vec{\sigma}\vec{q}~\langle B'|~\tilde{V}_\pi|B\rangle.
\end{align}
with $B,~B'$ the quark wave functions of Eq.~(\ref{wavefunc}).
Then we find the vertices,
\begin{align}\label{lowervet}
&    -it=\frac{4}{5}\frac{f_{\pi N}}{m_{\pi}}{\vec{\sigma}} ~ {\vec{q}}, \qquad\qquad \text{for~~ $\Sigma_c^{++} \pi^0\to\Sigma_c^{++}$},\\\nonumber
&    -it=\frac{4}{5}\frac{f_{\pi N}}{m_{\pi}}{\vec{\sigma}} ~ {\vec{q}}, \qquad\qquad \text{for~~ $\Sigma_c^{+} \pi^+\to\Sigma_c^{++}$},\\\nonumber
&    -it=-\frac{2\sqrt{3}}{5}\frac{f_{\pi N}}{m_{\pi}}{\vec{\sigma}} ~ {\vec{q}}, \qquad \text{for~~ $\Sigma_c^{++} \pi^-\to\Lambda_c^{+}$},\\\nonumber
&    -it=\frac{2\sqrt{3}}{5}\frac{f_{\pi N}}{m_{\pi}}{\vec{\sigma}} ~ {\vec{q}}, \qquad \text{for~~ $\Sigma_c^{+} \pi^0\to\Lambda_c^{+}$}.\\\nonumber
\end{align}

%With the inclusion of both types of vertices, we obtain the weights of each diagram shown in Fig.~\ref{feynDiag2}, which are given by:

%\begin{align}
%&    \text{FAC($P_c\to\bar{D}\Sigma_c$)}=\frac{2}{3}(-\frac{1}{2}g\frac{4}{5}\frac{f_{\pi N}}{m_{\pi}})-\frac{1}{3}(-g)\frac{4}{5}\frac{f_{\pi N}}{m_{\pi}}=-\frac{1}{3}g\frac{8}{5}\frac{f_{\pi N}}{m_{\pi}}, \qquad\qquad \text{for~~ $P_c\to\bar{D}\Sigma_c$}\\\nonumber
%&    \text{FAC($P_c\to\bar{D}\Lambda_c$)}=\frac{2}{3}g(-\frac{2\sqrt{3}}{5}\frac{f_{\pi N}}{m_{\pi}})-\frac{1}{3}\frac{1}{2}g\frac{2\sqrt{3}}{5}\frac{f_{\pi N}}{m_{\pi}}=g\frac{3\sqrt{2}}{5}\frac{f_{\pi N}}{m_{\pi}}, \qquad\qquad \text{for~~ $P_c\to\bar{D}\Lambda_c$}\\\nonumber
%\end{align}

{The structure of Eq.~(\ref{a}}) comes from an upper vertex of type $(p_\mu + p'_\mu)\vec{\epsilon}~\vec{\epsilon}^{~'} \tilde{\epsilon}^{\mu}$  and a lower vertex of the type 
$\gamma^\nu\tilde{\epsilon}_\nu $, with  $\tilde{\epsilon}^{\mu(\nu)}$,  the polarization of the exchanged vector. 
Both vertices select $\mu=0,~~\nu=0$ 
for small    momenta of the external particles. Hence for  small external three momenta  we get
\begin{align}
    \langle B' | \gamma^\mu | B \rangle (p_\mu + p'_\mu)\equiv (p^0+p^{'0})\delta_{mm'},
\end{align}
with $m,~m'$ the third components of the spin of $B,~B’$, respectively.
\subsection{Separation of spin $1/2$ and spin $3/2$}
To separate this amplitude $VB\to V'B'$ of Eq.~(\ref{a}) into contributions from the \(\frac{3}{2}\) and \(\frac{1}{2}\) states, we consider two structures. The first one is 
\begin{align}
{\vec{S}} \cdot {\vec{\epsilon}}{~'} ~ {\vec{S}}^+ \cdot {\vec{\epsilon}},
\end{align}
where \({\vec{S}}^+\) is the spin transition operator for the transition from spin \(\frac{1}{2}\) to \(\frac{3}{2}\), defined according to the  Wigner-Eckart theorem, as
\begin{equation}
\begin{aligned}
   \langle \frac{3}{2} \, M | &{S}^{+}_{\mu} | \frac{1}{2} \, m \rangle=  \mathcal{C} \,(\frac{1}{2} \, 1 \, \frac{3}{2};\, m \, \mu \, M)\langle \frac{3}{2} || {\vec{S}}^{+} || \frac{1}{2} \rangle,\\
%    &  \uparrow  spherical \, basis \,({\vec{S}}^+ = ({S}^{+}_{+1}, {S}^{+}_{0}, {S}^{+}_{-1})) 
\end{aligned}  
\end{equation}
with \({{S}}_\mu^+\) the component of \({\vec{S}}^+\) in spherical basis,
and $\langle \frac{3}{2} || {\vec{S}}^{+} || \frac{1}{2} \rangle$ (reduced matrix element)  defined as $\langle \frac{3}{2} || {\vec{S}}^{+} || \frac{1}{2} \rangle \equiv 1$.
%%%%%%%%%%%%%%%%%%%%%%%%%%%%%%%%%%%%%%%%%%%%%%%
%%%%%%%%%%%%%%%%%%%%%%%%%%%%%%%%%%%%%%%%%%%%%%%

\par In actual calculations one makes use of the relationship 
\begin{equation}\label{relationship}
\begin{aligned}
    \sum_{M}\langle m'| &{S}_{i} | M \rangle \langle M | {S}^{+}_{j} | m \rangle = \langle m' | \frac{2}{3}\delta_{ij}-\frac{i}{3}\epsilon_{ijk}{\sigma}_{k} | m \rangle.\\
%    &\uparrow cartesian \, indice
\end{aligned}
\end{equation}
%
%with ${\sigma}$ the ordinary spin Pauli matrices for spin $1/2$.
 The structure ${\vec{S}} \cdot {\vec{\epsilon}}{~'} ~ {\vec{S}}^+ \cdot {\vec{\epsilon}}$, by construction, projects over spin  $\frac{3}{2}$.
 There is another structure: ${\vec{\sigma}} \cdot {\vec{\epsilon}}{~'} ~ {\vec{\sigma}} \cdot {\vec{\epsilon}}$, which projects over spin $1/2$. By iterating these structures using Eq.~(\ref{relationship}) and
 \begin{equation}\label{24}
   \sum\limits_{\tilde{m}}\langle m'|\sigma_i|\tilde{m}\rangle 
   \langle \tilde{m}|\sigma_j|m\rangle =
   \langle m'|\delta_{ij}+i\epsilon_{ijk}\sigma_{k}|m\rangle 
\end{equation}
and  summing over intermediate polarizations ($\sum\limits_{pol}\epsilon_{i}\epsilon_{j}=\delta_{ij}$), we can see that the projectors, properly normalized such that $P_i^2=P_i$, are 
\begin{align}\label{spinprop}
& P^{(3/2)}={\vec{S}}\cdot {\vec{\epsilon}}{~'} ~ {\vec{S}^+} \cdot {\vec{\epsilon}},\\\nonumber
&P^{(1/2)}=\frac{1}{3}~{\vec{\sigma}} \cdot {\vec{\epsilon}}{~'}~ {\vec{\sigma}}\cdot {\vec{\epsilon}}.
\end{align}

This implies that the vertices $P_c\to \bar{D}^*\Sigma_c$ in Fig.~\ref{feynDiag2} can be expressed as
\begin{equation}
    \begin{aligned}
        &g\Vec{S}\,\Vec{\epsilon}, \quad\quad~~ \text{for}  \quad \frac{3}{2}^-,\\
        &\frac{1}{\sqrt{3}}g\Vec{\sigma}\,\Vec{\epsilon}, \quad \text{for} \quad \frac{1}{2}^-.
    \end{aligned}
\end{equation}
Counting now the Clebsch–Gordan coefficients $\sqrt{\frac{2}{3}}$ and $-\sqrt{\frac{1}{3}}$ of Fig.~\ref{feynDiag2} for the couplings of  $P_c$ to the ${D}^{*-} \Sigma_c^{++} $ and $ \bar{D}^{*0} \Sigma_c^{+} $ components, the coefficients appearing in Eq.~(\ref{uppervet}) for the upper vertices and Eq.~(\ref{lowervet}) for the lower vertices, we obtain a structure, up to the propagators of the particles, 
\begin{align}\label{oprator}
    &(-i) g_{P_c, \bar{D}^*\Sigma_c} (-i) \vec{\epsilon} \cdot (\vec{q} - \vec{p}_1)  \vec{\sigma} \cdot \vec{q} \, { \begin{Bmatrix}\vec{S}  \vec{\epsilon} \,\\\frac{1}{\sqrt{3}} \vec{\sigma}  \vec{\epsilon}\, \end{Bmatrix}}~ \mathrm{FAC},
\end{align}
with 
\begin{align}\label{factors}
 \text{FAC}=\left\{\begin{matrix}
 -\frac{1}{\sqrt{3}}\frac{8}{5}g\frac{f_{\pi N}}{m_{\pi}}, \qquad\qquad \text{for~~ $P_c\to {D}^-\Sigma_c^{++}$},\\
 -\frac{3\sqrt{2}}{5}g\frac{f_{\pi N}}{m_{\pi}}, \qquad\qquad \text{for~~ $P_c\to\bar{D}^0\Lambda_c^+$},
\end{matrix}\right.
\end{align}
where \{\} stands for $\left\{ 
         \begin{aligned}
           & 3/2 \\
           & 1/2 \\
         \end{aligned} \right\}$.

The loop function for the triangle diagrams of Fig.~\ref{feynDiag2} is then given by
\begin{equation}
\begin{aligned}
    -it=&~ \text{FAC}~\int \frac{d^4q}{(2\pi)^4} (-i)g_{P_c,\bar{D}^*\Sigma_c}  (-i) \vec{\epsilon} \cdot(\vec{q} - \vec{p}_1)~  \vec{\sigma} \cdot \vec{q}~ { \begin{Bmatrix}\vec{S}  \vec{\epsilon} \,\\\frac{1}{\sqrt{3}} \vec{\sigma}  \vec{\epsilon}\, \end{Bmatrix}}\\
    &\times \frac{i}{q^2 - m_\pi^2 + i\varepsilon}\frac{i}{(q + p_1)^2 - m_{D^*}^2 + i\varepsilon}\frac{i2 M_{\Sigma_c}}{2 E_{\Sigma_c} (\vec{q} + \vec{p}_1)}\frac{1}{P^0 - q^0 - {p_1}^0 - E_{\Sigma_c}(\vec{q} + \vec{p}_1) + i\varepsilon}.
\end{aligned}
\end{equation}
Then we separate the positive and negative components of the $D^*$  propagator as 
\begin{equation}
    \begin{aligned}
    \frac{1}{(q + p_1)^2 - m_{D^*}^2 + i\varepsilon}
    &=\frac{1}{2 \omega_{D^*}(\vec{q} + \vec{p}_1)}\left(\frac{1}{q_{0} + p_{1}^{0} - \omega_{D^*}(\vec{q} + \vec{p}_{1}) + i \varepsilon}-\frac{1}{q_{0} + p_{1}^{0} + \omega_{D^*}(\vec{q} + \vec{p}_{1}) - i \varepsilon}\right),
\end{aligned}
\end{equation}
disclosing the positive and negative energy parts of the propagator, and because the mass of the $\bar{D}^*$ is large and it is close to on shell in the loop, we keep only the positive energy part,
\begin{equation}
    \begin{aligned}
\frac{1}{2 \omega_{D^*}(\vec{q} + \vec{p}_1)}\frac{1}{q_{0} + p_{1}^{0} - \omega_{D^*}(\vec{q} + \vec{p}_{1}) + i \varepsilon}.
\end{aligned}
\end{equation}
%$p_1^0$ represents the $D$ momentum, We  multiply the  $ \pi $ propagator by a form factor, $ F(q) = \frac{\Lambda^2}{q^2+\Lambda^{2}}  $ and  $ F(q) = e^{-q^2/\Lambda^{2} } $, where $  \Lambda =1000$~\text{MeV}. 
Similarly, for the pion propagator we have
%%%%%%%%%%%%%%%%%%%%%%%%%%%%%%%%%%%%%%%%%%%%%%%
%For the pion exchange we take the full propagation:
\begin{equation}
    \frac{1}{q^2-m_\pi^2}=\frac{1}{2\omega_\pi(\vec{q}\,)}\left(\frac{1}{q^0 - \omega_\pi(\vec{q}\,) + i \varepsilon} - \frac{1}{q^0 + \omega_\pi(\vec{q}\,) - i \varepsilon}\right),
\end{equation}
%%%%%%%%%%%%%%%%%%%%%%%%%%%%%%%%%%%%%%%%%%%%%%%
%We take the positive energy part of the heavy, and close to on shell, $D^*$ propagators but keep the full $\pi$ propagator which will be off shell.
and because it is quite off shell, we keep the two terms.
Next we sum over the $D^*$ polarizations ($\sum\limits_{\text{pol}} \epsilon_{i} \epsilon_{j} = \delta_{ij}$),

\begin{equation}
   \sum\limits_{\text{pol}} \epsilon_{i} \left(q - p_{1}\right)_{i}
\begin{Bmatrix}S_{j} \epsilon_{j}\\\frac{1}{\sqrt{3}} \sigma_{j} \epsilon_{j}
\end{Bmatrix}=\begin{Bmatrix}\left(q - p_{1}\right)_{j}S_{j}\\\frac{1}{\sqrt{3}}\left(q - p_{1}\right)_{j} \sigma_{j}
\end{Bmatrix}.
\end{equation}

%
%\begin{equation}
%\sum_{\text{pol}} \epsilon_{i} \epsilon_{j} = \delta_{ij}
%\end{equation}
%
%\begin{equation}
%\sum_{\text{pol}}  \epsilon_{i} \left(q - p_{1}\right)_{i}S_{j}\epsilon_{j} = \delta_{ij}\left(q - p_{1}\right)_{i}S_{j}=\left(q - p_{1}\right)_{j}S_{j}
%\end{equation}
%
%\begin{equation}
%    \epsilon_{i} \left(q - p_{1}\right)_{i}
%\begin{Bmatrix}S_{j} \epsilon_{j}\\\frac{1}{\sqrt{3}} \sigma_{j} \epsilon_{j}
%\end{Bmatrix}=\begin{Bmatrix}\left(q - p_{1}\right)_{j}S_{j}\\\frac{1}{\sqrt{3}}\left(q - p_{1}\right)_{j} \sigma_{j}
%\end{Bmatrix}
%\end{equation}
%
Thus we have
\begin{equation}\label{allt}
    \begin{aligned}
    -it=&i~ \text{FAC}~ \,\,g_{P_c,\bar{D}^*\Sigma_c}\int \frac{d^4q}{(2\pi)^4}\, \vec{\sigma} \cdot \vec{q}\,  \begin{Bmatrix}\left(q - p_{1}\right)_{j}S_{j}\\\frac{1}{\sqrt{3}}\left(q - p_{1}\right)_{j} \sigma_{j}\end{Bmatrix} \frac{2 M_{\Sigma_c}}{2 E_{\Sigma_c} (\vec{q} + \vec{p}_1)} \frac{1}{P^0 - q^0 - {p_1}^0 - E_{\Sigma_c}(\vec{q} + \vec{p}_1) + i\varepsilon}\\
    & \times \frac{1}{2 \omega_{D^*}(\vec{q} + \vec{p}_1)}  \frac{1}{q^{0} + p_{1}^{0} - \omega_{D^*}(\vec{q} + \vec{p}_{1}) + i \varepsilon} \frac{1}{2\omega_\pi(\vec{q}\,)}\left(\frac{1}{q^0 - \omega_\pi(\vec{q}\,) + i \varepsilon} - \frac{1}{q^0 + \omega_\pi(\vec{q}\,) - i \varepsilon}\right).
\end{aligned}
\end{equation}
%%%%%%%%%%%%%%%%%%%%%%%%%%%%%%%%%%%%%%%%%%%%%%%
The \(q^0\) integration is performed using Cauchy's theorem, applying the residue method for evaluation, and finally, we find 
    \begin{align}\label{35}\nonumber
    -    it=&\text{FAC}\,g_{P_c,\bar{D}^*\Sigma_c}\int \frac{d^3q}{(2\pi)^2}\, \sigma_{i}\,q_{i}\begin{Bmatrix}\left(q - p_{1}\right)_{j}S_{j}\\
    \frac{1}{\sqrt{3}}\left(q - p_{1}\right)_{j} \sigma_{j}\end{Bmatrix} \frac{2 M_{\Sigma_c}}{2 E_{\Sigma_c} (\vec{q} + \vec{p}_1)}\frac{1}{2 \omega_{D^*}(\vec{q} + \vec{p}_1)}\frac{1}{2\omega_\pi(\vec{q}\,)}\\
    & \times \frac{1}{P^0 - E_{\Sigma_c}(\vec{q} + \vec{p}_1)- \omega_{D^*}(\vec{q} + \vec{p}_{1})+i\varepsilon}
    \left(\frac{1}{P^0 -p_{1}^{0}- E_{\Sigma_c}(\vec{q} + \vec{p}_1)- \omega_\pi(\vec{q}\,)+i\varepsilon}+\frac{1}{p_{1}^{0} - \omega_{D^*}(\vec{q} + \vec{p}_{1})-\omega_\pi(\vec{q}\,)+i\varepsilon}\right).
\end{align}
%
%With a bit of algebra, using Eqs.~(\ref{relationship}) ~(\ref{24}) and the results of Appendix A
%Equation of $\sigma_i\sigma_j\to\delta_{ij}+i\epsilon_{ijk}$,
%we can write the transition amplitude \( t \) for Fig.~\ref{feynDiag2}  as:
%
In Eq.~(\ref{35}) we find two structures, one linear in  $\vec{q}$, and another one quadratic in $q$. We can write
\begin{align}
   & \int\frac{d^3~q}{(2\pi)^3}~f(\Vec{q},\,\Vec{p}_1)q_i=a~ p_{1i},  \\\nonumber
     & \int\frac{d^3~q}{(2\pi)^3}~f(\Vec{q},\,\Vec{p}_1)q_iq_j=b~ {\vec{p}_1^{~2}}\delta_{ij}+c~p_{1i}~p_{1j}.
\end{align}

%\begin{align}\label{tmatrix}
%& -i t=\frac{1}{\sqrt{3}}(c-a+3 b) p_1^2 \quad \text { for } 1 / 2^- \\\nonumber
%&  {-it }=\sqrt{\frac{1}{3}}(c-a) p_1^2 \quad \quad \quad ~\text { for } 3 / 2^- 
%\end{align}
%
Details on how to calculate $a,~b,~c$ are given in Appendix A. As a consequence we obtain
\begin{align}\label{tmatrix}
 -i t=& -a~ \sigma_ip_{1i}  \begin{Bmatrix} p_{1j}S_{j}\\
    \frac{1}{\sqrt{3}} p_{1j} \sigma_{j}\end{Bmatrix}
 + b~ \vec{p}_1^{~2} \sigma_i \begin{Bmatrix} S_{i}\\
    \frac{1}{\sqrt{3}} \sigma_{i}\end{Bmatrix} 
 + c~\sigma_ip_{1i}  \begin{Bmatrix} p_{1j}S_{j}\\
    \frac{1}{\sqrt{3}} p_{1j} \sigma_{j}\end{Bmatrix} .
\end{align}
It is easy to see that $\sigma_i~S_i$   in Eq.~(\ref{tmatrix}) is a scalar which cannot make transition from spin $3/2$ to $1/2$ and hence is zero, such that Eq.~(\ref{tmatrix})  {can be recast using} Eq.~(\ref{factors})   as
\begin{align}\label{28_1}
   \begin{Bmatrix} \langle m'|(c-a)\sigma_{i}S_{j} p_{1i} p_{1j}|M\rangle\\
    \langle m'|\frac{1}{\sqrt{3}}(c-a+3b)\vec{p}_1^{~2}|m\rangle \end{Bmatrix} ,
\end{align}
with $m'$ the final $\Sigma_c~(\Lambda_c)$  spin and $M,~m$ the initial spin for $J=3/2,~1/2$ respectively. Using now Eq.~(\ref{relationship}) we can do the sum and average of $|t|^2$ for each case,
$\frac{1}{4}\sum\limits_{M,m'}|t|^2$  for $J=3/2$ and  $\frac{1}{2}\sum\limits_{M,m'}|t|^2$  for $J=1/2$  and we immediately find 
\begin{align}\label{T2}
\bar{\sum}\sum |t|^2=
   \begin{Bmatrix}{\frac{1}{3}}~|c-a|^2~ \vec{p}_1^{~4} \\\frac{1}{{3}}~ |c-a+3 b|^2~ \vec{p}_1^{~4} 
   \end{Bmatrix} .
\end{align}
It is also easy to show, as it should be, that from the contribution $p_{1i} p_{1j}\equiv(p_{1i} p_{1j}-\frac{1}{3}\vec{p}_1^{~2}\delta_{ij})+\frac{1}{3}\vec{p}_1^{~2}\delta_{ij}$, which separates the $D-$wave part of the  $\bar{D}\Sigma_c~(\bar{D}\Lambda_c)$  final state from the $S-$
wave part, only the $S-$wave part contributes for  $1/2^-$ and only the $D-$wave part contribute for  $3/2^-$.

The magnitudes  $a,~b,~c$  calculated in Appendix A are given by
\begin{align}\label{abc}
&  a=\int \frac{d^3 q}{(2 \pi)^3} \frac{\vec{q} \cdot \vec{p}_1}{\vec{p}_1^{~2}}~  A ~ \Theta\left(q_{\max }-\left|\vec{p}_1+\vec{q}\,\right|\right) ~ FF(q) ,\\\nonumber
&  b=\int \frac{d^3 q}{(2 \pi)^3} \frac{1}{2 \vec{p}_1^{~4}}\left[\vec{q}^{~2} \vec{p}_1^{~2}-\left(\vec{q} \cdot \vec{p}_1\right)^2\right]~   A ~ \Theta\left(q_{\max }-\left|\vec{p}_1+\vec{q}\,\right|\right)  ~ FF(q), \\\nonumber
& c=\int \frac{d^3 q}{(2 \pi)^3} \frac{1}{2 \vec{p}_1^{~4}}\left[3\left(\vec{q} \cdot \vec{p}_1\right)^2-\vec{q}^{~2} \vec{p}_1^{~2}\right]~   A ~ \Theta\left(q_{\max }-\left|\vec{p}_1+\vec{q}\,\right|\right) ~ FF(q),
\end{align}
with the factor $A$  defined by
\begin{align}
    A=&\,\,\text{FAC}\, g_{P_c,\bar{D}^*\Sigma_c} \frac{2M_{\Sigma_c}}{2E_{\Sigma_c}(\vec{q}+\vec{p}_1)} \frac{1}{2\omega_{D^*}(\vec{q}+\vec{p}_1)} \frac{1}{2\omega_\pi(\vec{q})}
\frac{1}{P^0-\omega_{D^*}\left(\vec{q}+\vec{p}_1\right)-E_{\Sigma_c}\left(\vec{q}+\vec{p}_1\right)+i \varepsilon}  \\\nonumber
& \times\left[\frac{1}{ P^0-p_1^0-E_{\Sigma_c}\left(\vec{q}+\vec{p}_1\right)-\omega_\pi(\vec{q}\,)+i \varepsilon}+\frac{1}{p_1^0-\omega_{D^*}\left(\vec{q}+\vec{p}_1\right)-\omega_\pi(\vec{q}\,)+i \varepsilon}\right]. 
\end{align}
with 
\begin{align}
& P^0=M_{P_c},\qquad\qquad p_1^0=\frac{M_{P_c}^2+m_D^2-M_B^2}{2 M_{P_c}} ,
\end{align}
with $M_B$ corresponding to  $M_{\Sigma_c}$ and $M_{\Lambda_c}$.
%The amplitudes in Eq.~(\ref{tmatrix}) should be used without any reference to spin, since sum and average over spins in $|t|^2$ are already implicit in this form.
\par In  Eq.~(\ref{abc}) we have added the regulators of the loop. On the one hand, we add the factor $\Theta(q_\text{max}-|\vec{p}_1+\vec{q}\,|)$ because this factor appears when we regularize the $G$ function of {Eq.~(\ref{Gfun}) with this cut-off factor. This implies that the $T$-matrix stemming {from Eq.~(\ref{c})} is of the form~\cite{Gamermann:2009uq}
\begin{align}
 T\left(\vec{p}^{~\prime}, \vec{p}\,\right)= \Theta\left(q_{\max }-|\vec{p}~|\right) \Theta\left(q_{\max }-|\vec{p}~{'}|\right)T ,  
\end{align}
and hence, the $P_c\to\bar{D}^*({q}+{p}_1)\Sigma_c({P}-{q}-{p}_1)$ vertex contains the factor $\Theta\left(q_{\max }-|\vec{q}+\vec{p}_1|\right)$. On the other hand, it is customary for  off shell pion exchange to use an extra form factor. We use here
\begin{align}\qquad FF(q) = e^{-\vec{q}^{~2}/\Lambda^{2}}.
\end{align}
%and we take $\Lambda=1$~GeV similar as to what was used in~\cite{Molina:2020hde}, but this $FF(q)$ has very moderate effects once the cut off $\Theta\left(q_{\max }-|\vec{q}+\vec{p}_1|\right)$ is used.

In~\cite{Molina:2020hde} values of $\Lambda\approx1000$~MeV  were used. 
We take here $\Lambda \approx 950$~MeV, by means of which we obtain the largest width around 20.5~MeV, close to the experiment. We note that if we take $\Lambda\approx1000$~MeV, the results change moderately and this number becomes 22.2~MeV. It is clear that the regulator  $\Theta(q_\text{max}-|\vec{p}_1+\vec{q}\,|)$  is already quite effective, stabilizing the $\Lambda$ dependence.
Then, we can calculate the width for each diagram as
\begin{align}\label{widthtotal}
& \Gamma=\frac{2 M_B 2 M_{P_c}}{8 \pi}\frac{1}{M_{P_c}^2}|t|^2 p_1  =\frac{1}{2 \pi} \frac{M_B}{M_{P_c}}|t|^2 p_1 ,
\end{align}
where $p_1$ is defined as
\begin{align}
& p_1=\frac{\lambda^{1 / 2}\left(M_{P_c}^2,~ m_D^2,~ M_B^2\right)}{2 M_{P_c}}.
\end{align}

It is also easy to see that we should expect a larger contribution for $1/2^-$ than $3/2^-$. Indeed, in the limit of small $|\vec{p}_1|$ we can make the substitution $q_iq_j\to \frac{1}{3}\vec{q}^{~2}\delta_{ij}$   in Eq.~(\ref{abc}), which makes $c\to 0$,  also $a$ would go to zero, but not $b$, which only affects the $1/2^-$ state. The actual case is not so extreme and one gets a contribution for both $1/2^-$ and $3/2^-$   but the one for $1/2^-$   is bigger. 
Finally, a small detail. We have evaluated the $P_c$  decay to ${D}^-\Sigma_c^{++}$. To account for the extra $\bar{D}^0\Sigma_c^{+}$ we simply have to take into  account the    Clebsch–Gordan coefficients of Eq.~(\ref{1}) ($\bar{D}$ has the same isospin as $\bar{D}^*$) and the decay width to $\bar{D}^0\Sigma_c^{+}$   is $\frac{1}{2}$ of that to ${D}^-\Sigma_c^{++}$. Then to account for all $\bar{D}\Sigma_c$ decay, we multiply the results of Eq.~(\ref{widthtotal})  by $\frac{3}{2}$. The results that follow refer to the total $\bar{D}\Sigma_c$  decay width.
%\newpage
\section{results and discussions}
\subsection{$P_c \to \bar{D}\Sigma_c,~\bar{D}\Lambda_c$ widths}

We select the masses and widths from $P_c\to J/\psi N$ decay and couplings corresponding to $q_\mathrm{max}=450$~MeV and $q_\mathrm{max}=580$~MeV from Table~\ref{POLE_LIANG}. 
%To estimate uncertainties, we employed two different forms of form factors: $FF(q) = \frac{\Lambda^2}{\Vec{q}^2+\Lambda^{2}}$ and $FF(q) = e^{-\Vec{q}^2/\Lambda^{2}}$. 
Then, we calculate the decay widths of $P_c \to \bar{D}\Sigma_c$ with $J^P=1/2^-,~~J^P=3/2^-$, and $P_c \to \bar{D}\Lambda_c$ with $J^P=1/2^-,~~J^P=3/2^-$. The results are provided in Tables ~\ref{form2_1} and ~\ref{form2_2}.

\begin{table}[H]
%\footnotesize
\centering
\caption{Decay widths for the $P_c(4457)$   assuming different spins (in units of MeV).}
\label{form2_1}
\setlength{\tabcolsep}{38pt}
\begin{tabular}{ccc}
\hline \hline
$ J^P$ &$\frac{1}{2}^-$ & $ \frac{3}{2}^-$ \\
\hline
  \multirow{3}*{$P_c(4457)$}~~$\bar{D} \Sigma_c$ & 5.60  & 4.11  \\ 
\phantom{ \multirow{2}*{$P_c(4457)$}~~}$\bar{D} \Lambda_c$ &  2.47  &  2.15  \\
\phantom{ \multirow{2}*{$P_c(4457)$}~~}$J/ \psi N$ & 1.52  & 1.52  \\
\hline
\end{tabular}
\end{table}

\begin{table}[H]
%\footnotesize
\centering
\caption{Decay widths for the $P_c(4440)$   assuming different spins (in units of MeV).}
\label{form2_2}
\setlength{\tabcolsep}{38pt}
\begin{tabular}{ccc}
\hline \hline
$ J^P$ &$\frac{1}{2}^-$ & $ \frac{3}{2}^-$ \\
\hline
  \multirow{3}*{$P_c(4440)$}~~$\bar{D} \Sigma_c$ & 10.43   & 5.35  \\ 
\phantom{ \multirow{2}*{$P_c(4440)$}~~}$\bar{D} \Lambda_c$ &  5.98  &  4.53   \\
\phantom{ \multirow{2}*{$P_c(4457)$}~~}$J/ \psi N$ &  4.10  &  4.10   \\
\hline
\end{tabular}
\end{table}

%Based on the width information in Table~1, we determined the decay widths of $P_c \to \bar{D}\Sigma_c$ and $P_c \to \bar{D}\Lambda_c$ for $J^P=1/2^-$ and $J^P=3/2^-$. These are summarized in Table~~\ref{form_total}. 

%Wee take average values and dispassion from the results in Tables~\ref{form1_1},  ~\ref{form2_1}, \ref{form1_2}, and ~\ref{form2_2} summing the three sources of widths, and 
Next  we look at two possible scenarios: one where the $P_c(4440)$, $P_c(4457)$   correspond to $1/2^-,~~3/2^-$, or viceversa. The results are shown in Table~~\ref{form_total}.

\begin{table}[H]
%\footnotesize
\centering
\caption{ Total Width of $ P_c\to \bar{D} \Sigma_c,~~\bar{D}\Lambda_c$ decay (in units of MeV).}
\label{form_total}
\setlength{\tabcolsep}{38pt}
\begin{tabular}{ccc}
\hline \hline 
\multirow{2}*{ \textbf{Scenario 1}}   & $P_c(4440) ~(J^P=\frac{1}{2}^-)$  & $P_c(4457) ~(J^P=\frac{3}{2}^-)$ \\
& 20.52   & 7.78  \\
\multirow{2}*{ \textbf{Scenario 2}}& $P_c(4440) ~(J^P=\frac{3}{2}^-)$  & $P_c(4457) ~(J^P=\frac{1}{2}^-)$ \\ 
& 13.98   &  9.59   \\
\multirow{2}*{ \text{Exp.}}& $P_c(4440)$  & $P_c(4457)$ \\ 
& $20.6 \pm 4.9_{-10.1}^{+8.7} $  &  $6.4 \pm 2.0_{-1.9}^{+5.7}$  \\
\hline
\end{tabular}
\end{table}
%%%%%%%%%%%%%%%%%
%%%%%%%%%%%%%%%%%
By analyzing the data, we conclude:\\  
\textbf{Scenario 1:} If $P_c(4440)$ has quantum numbers $J^P=1/2^-$ and $P_c(4457)$ has $J^P=3/2^-$, their widths are $20.5$~MeV and $7.8$~~\text{MeV}, respectively.  \\
\textbf{Scenario 2:} If $P_c(4440)$ corresponds to $J^P=3/2^-$ and $P_c(4457)$ to $J^P=1/2^-$, the widths are $14.0$~MeV and $9.6$~\text{MeV}, respectively.  

When comparing these outcomes with experimental observations~\cite{LHCb:2019kea}, we find that both Scenarios are compatible with the experimental data within errors.
Yet, if we look at the ratio of the central values for the widths of $P_c(4440)$ to $P_c(4457)$, we find a ratio 2.64 for the first scenario and 1.46 for the second scenario, compared with the experimental ratio 3.22. It is clear that the scenario 1 ($1/2^-,~~3/2^-$)  is preferable.
This suggests that $P_c(4440)$ likely corresponds to $J^P=1/2^-$, and $P_c(4457)$ is likely associated with $J^P=3/2^-$. 

Next, we separately analyze the decay widths of $P_c \to \bar{D}\Sigma_c$ and $P_c \to \bar{D}\Lambda_c$ for various quantum numbers. The  results are summarized in Tables~\ref{decay_1} and \ref{decay_2}, respectively.

\begin{table}[H]
%\footnotesize
\centering
\caption{Width of $P_c\to  \bar{D} \Sigma_c$ (in units of MeV).}
\label{decay_1}
\setlength{\tabcolsep}{38pt}
\begin{tabular}{ccc}
\hline \hline 
\multirow{2}*{ \textbf{Scenario 1}}    & $P_c(4440) ~~(J^P=\frac{1}{2}^-)$ & $P_c(4457) ~(J^P=\frac{3}{2}^-)$ \\
 &  10.43  &  4.11 \\
\multirow{2}*{ \textbf{Scenario 2}}  & $P_c(4440) ~(J^P=\frac{3}{2}^-)$ & $P_c(4457) ~(J^P=\frac{1}{2}^-)$ \\
 &  5.35 &  5.60  \\
\hline
\end{tabular}
\end{table}
%%%%%%%%%%%%%%%%%%%%%%%%%%%%%%%%%%%%%%%%%%%%%%%
%%%%%%%%%%%%%%%%%%%%%%%%%%%%%%%%%%%%%%%%%%%%%%%

\begin{table}[H]
%\footnotesize
\centering
\caption{Width of $P_c\to  \bar{D} \Lambda_c$ (in units of MeV).}
\label{decay_2}
\setlength{\tabcolsep}{38pt}
\begin{tabular}{ccc}
\hline \hline 
\multirow{2}*{ \textbf{Scenario 1}}   & $P_c(4440) ~~(J^P=\frac{1}{2}^-)$ & $P_c(4457) ~(J^P=\frac{3}{2}^-)$ \\
 &  5.98  &  2.15\\
\multirow{2}*{ \textbf{Scenario 2}} & $P_c(4440) ~(J^P=\frac{3}{2}^-)$  & $P_c(4457) ~(J^P=\frac{1}{2}^-)$ \\
&   4.53 &   2.47 \\
\hline
\end{tabular}
\end{table}
For the $P_c \to \bar{D}\Sigma_c$ decay widths with $J^P=1/2^-$ and $J^P=3/2^-$ (Table~\ref{decay_1}), we find:  \\
\textbf{Scenario 1:} Assuming $P_c(4440)$ corresponds to $J^P=1/2^-$ and $P_c(4457)$ to $J^P=3/2^-$, the respective widths are $10.4$~MeV and $4.1$~~\text{MeV}, leading to the ratio:  
\[
\frac{\Gamma_{P_c(4440) \to \bar{D}\Sigma_c}}{\Gamma_{P_c(4457) \to \bar{D}\Sigma_c }} \approx 2.5.
\]  \\
\textbf{Scenario 2:} If $P_c(4440)$ has $J^P=3/2^-$ and $P_c(4457)$ has $J^P=1/2^-$, the widths are $5.4$~MeV and   $5.6$~\text{MeV}, resulting in the ratio:  
\[
\frac{\Gamma_{P_c(4440) \to \bar{D}\Sigma_c}}{\Gamma_{P_c(4457) \to \bar{D}\Sigma_c }} \approx 0.96.
\]  
Thus, if future experimental results show that the $P_c \to \bar{D}\Sigma_c$ width with $P_c(4440)$ is approximately 2.5 times larger than that with $P_c(4457)$, it would support the hypothesis that $P_c(4440)$ corresponds to $J^P=1/2^-$ and $P_c(4457)$ to $J^P=3/2^-$. Conversely, if the widths are nearly equal, it would imply that $P_c(4440)$ corresponds to $J^P=3/2^-$ and $P_c(4457)$ to $J^P=1/2^-$.  

For the  $P_c \to \bar{D}\Lambda_c$ decay widths with $J^P=1/2^-$ and $J^P=3/2^-$ (Table~\ref{decay_2}), we observe: \\ 
\textbf{Scenario 1:} If $P_c(4440)$ has $J^P=1/2^-$ and $P_c(4457)$ has $J^P=3/2^-$, their widths are $6.0$~MeV and $2.2$~~\text{MeV}, yielding the ratio:  
\[
\frac{\Gamma_{P_c(4440) \to \bar{D}\Lambda_c}}{\Gamma_{P_c(4457) \to \bar{D}\Lambda_c}} \approx 2.8.
\]  \\
\textbf{Scenario 2:} Assuming $P_c(4440)$ corresponds to $J^P=3/2^-$ and $P_c(4457)$ to $J^P=1/2^-$, the widths are $4.5$~MeV and $2.5$~~\text{MeV}, respectively, leading to the ratio:  
\[
\frac{\Gamma_{P_c(4440) \to \bar{D}\Lambda_c}}{\Gamma_{P_c(4457) \to \bar{D}\Lambda_c}} \approx 1.8.
\]  
Therefore, if future experiments determine that the $P_c(4440) \to \bar{D}\Lambda_c$  width is approximately three times larger than the $P_c(4457)$ width, it would favor $P_c(4440)$ having $J^P=1/2^-$ and $P_c(4457)$ having $J^P=3/2^-$. On the other hand, if the ratio of widths is about 2, it would indicate that $P_c(4440)$ corresponds to $J^P=3/2^-$ and $P_c(4457)$ to $J^P=1/2^-$.

Both  informations on the decay width      are, thus, instructive on the possible spin assignment, particularly the   decay  into $\bar{D}\Sigma_c$ in which these two scenarios produce drastic different ratios of widths.

%\newpage
\subsection{Mass splitting of the $P_c$ states }
We address now the possible mass splitting of the states due to the mechanism of $\pi$ exchange. For this we write the selfenergy of the state depicted in Fig.~\ref{figlast} as
\begin{figure}[h]
    \centering
    \includegraphics[width=0.6\textwidth]{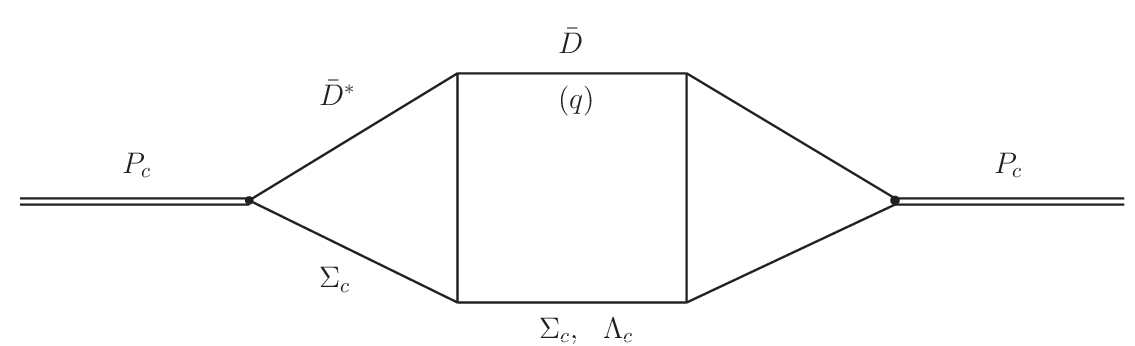}
    \caption{$P_c$ selfenergy through $\bar{D}\Sigma_{c},~\bar{D}\Lambda_{c}$    intermediate states.}
    \label{figlast}
\end{figure}

\begin{equation}
\begin{aligned}
        \Sigma_{\Sigma_c}&={|t_{P_c,\bar{D}\Sigma_{c}}|}^{2} \int \frac{d^3q}{(2\pi)^3}\,\frac{1}{2\omega_{D}(q)}\,\frac{M_{\Sigma_c}}{E_{\Sigma_c}(q)}\,\frac{1}{P^0-\omega_{D}(q)-E_{\Sigma_c}(q)+i\,\epsilon}\\&={|t_{P_c,\bar{D}\Sigma_{c}}|}^{2}G_{\bar{D}\Sigma_{c}}(P^0)
\end{aligned}
\end{equation}
and similarly
\begin{equation}
    \Sigma_{\Lambda_c}={|t_{P_c,\bar{D}\Lambda_{c}}|}^{2}G_{\bar{D}\Lambda_{c}}(P^0),
\end{equation}
with ${|t_{P_c,\bar{D}\Sigma_{c}}|}^{2}$, ${|t_{P_c,\bar{D}\Lambda{c}}|}^{2}$ given by Eq.~(\ref{T2}),
where G is the loop function for intermediate meson-baryon propagation used in the study of meson-baryon interaction.
%%%%%%%%%%%%%%%%%%%%%%%%%%%%%%%%%%%%%%%%%%%%%%%

The $\bar{D}^{*}\Sigma_{c}\to \bar{D}^{*}\Sigma_{c}$ amplitude will now be written by:
\begin{equation}\label{d}
t^{(j)}_{\bar{D}^{*}\Sigma_{c},\bar{D}^{*}\Sigma_{c}}=\,P^{(j)}\,\frac{g_{P_{c},{\bar{D}^*}\Sigma_{c}}^{2}}{\sqrt{s}-M_{P_c}+\frac{i\,\Gamma}{2}-\Sigma_{\Sigma_c}^{(j)}-\Sigma_{\Lambda_c}^{(j)}}, \qquad j=1/2^{-},~3/2^{-},
\end{equation}
 with $P^{(j)}$ the spin projector of { Eq.~(\ref{spinprop})}.
 and  $\Gamma/2$ from Table~\ref{POLE_LIANG} coming from $P_c\to J/\psi N$.
\par It is easy to see, evaluating $\text{Im\,}\Sigma_{\Sigma_c}$, $\text{Im\,}\Sigma_{\Lambda_c}$ from $\text{Im\,}G$, that $\text{Im\,}\Sigma_{\Sigma_c}=\frac{\Sigma_{\bar{D}\Sigma_c}}{2}$, $\text{Im\,}\Sigma_{\Lambda_c}=\frac{\Sigma_{\bar{D}\Lambda_c}}{2}$ as it should be. We take now a cut-off for $\bar{D}\Sigma_c$, $\bar{D}\Lambda_c$ of the order of 1 GeV, to be safely away from having $|\vec{q}\,|$ in Fig.~\ref{figlast} on shell for $P_c\to \bar{D}(\vec{q}\,)\Sigma_{c},\,\bar{D}(\vec{q}\,)\Lambda_{c}$ larger than $q_\mathrm{max}$, and evaluate also the real part of the amplitude of Eq.~(\ref{d}). We look at  ${|t^{(j)}_{\bar{D}^{*}\Sigma_{c},\bar{D}^{*}\Sigma_{c}}|}^2$ to see the position  of the peak and hence the new mass, and evaluate the width from the width of ${|t_{\bar{D}^{*}\Sigma_{c},\bar{D}^{*}\Sigma_{c}}|}^2$ as a function of $\sqrt{s}$. We start from $\sqrt{s}=4457\,$~MeV, assuming that both spin states originate from the same energy and we find the new masses and widths as:
\begin{equation}
    \begin{aligned}
        \Tilde{M}&=4454.82\,~\text{MeV},\qquad \Gamma= 9.63 \,~\text{MeV},\qquad(1/2^-)\\
        \Tilde{M}&=4455.39\,~\text{MeV},\qquad \Gamma= 7.81\,~\text{MeV},\qquad(3/2^-)
    \end{aligned}
\end{equation}
%%%%%%%%%%%%%%%%%%%%%%%%%%%%%%%%%%%%%%%%%%%%%%%

We can see a small splitting, driving the $1/2^-$ state to smaller energy than the $3/2^-$ state, but very small, insufficient to cope for the 17 MeV experimental splitting. We have taken here the $t_{P_c,\bar{D}\Sigma_{c}}$ amplitude on shell and one might wonder if the off shell extrapolation of this amplitude could induce larger splitting, but one would have to assume uncertainties due to the off shell extrapolation. Certainly one cannot rule out other reasons for the splitting, and the results obtained here open a new window of research. Yet, whatever the ultimate reason for it , what matters for the evaluation of the width is the couplings of $P_c$ to $\bar{D}^*\Sigma_c$ at the energies where the $P_c(4440)$ and $P_c(4457)$ are observed, which are detailed in Table~\ref{POLE_LIANG}, where the masses are obtained with the {two channel problem } varying the parameter $q_\mathrm{max}$. We should note that the effect of missing channels that can collaborate to the splitting of the mass can be incorporated via an effective potential in the remaining channels~\cite{Hyodo:2013nka,Aceti:2014ala}, and that changes in the potential can be trade off with changes in $q_\mathrm{max}$.
%%%%%%%%%%%%%%%%%%%%%%%
\par In the work of Refs.~\cite{Du:2019pij,Du:2021fmf} the splitting of the states is obtained using as input contact terms for the interaction of coupled channels, guided by heavy quark spin symmetry, plus pion exchange. If one looks into Ref.~\cite{Xiao:2013yca}, one can see that there are 7 free
 parameters for the interaction, plus other parameters to regularize the loops. With so much freedom it is also not easy to make definite conclusions concerning the spin of the states\footnote{We acknowledge useful discussions with Vadim Baru concerning this issue.}. The investigation of the widths done here has used only one parameter, fine tuning $\Lambda$ in the  pion exchange form factor, to get the precise strength of the largest width. All the other widths evaluated are predictions of the approach. We think that the precise measurement of these widths will be decisive to settle the issue of  the spin assignment to these two $P_c$ states.

%%%%%%%%%%%%%%%%%%%%%%%

\section{Conclusions} 
  We have addressed here the issue of the widths of the $P_c(4440)$ and $P_c(4457)$ pentaquark states. Given the broad consensus on the nature of these states as molecular states of $\bar{D}^* \Sigma_c$, we have started by recalculating these states with the channels $\bar{D}^* \Sigma_c$  and $J/\psi N$ as coupled channels within a unitary approach with the interaction obtained from an extension of the local hidden gauge approach exchanging vector mesons. By changing the regulator of the loops, $q_\mathrm{max}$, we reproduce the two masses and calculate the decay width to $J/\psi N$ for each mass. We see that this source of width is insufficient to obtain the experimental widths and then calculate explicitly the decay width to $\bar{D} \Sigma_c$ and $\bar{D}\Lambda_c$, driven by pion exchange. The pion exchange breaks the degeneracy in spin-parity  $1/2^-$ and $3/2^-$ of the vector exchange. We introduce some suitable projectors over the different spins and obtain reasonable values of the total widths compared with experiment. Then we look for possible assignments of the widths obtained to the 
  $P_c(4440)$ and $P_c(4457)$ states and show a preference of the results for the $1/2^-$ , $3/2^-$ order of the states.  The separate calculation  of the widths for the 
  $J/\psi N$, $\bar{D} \Sigma_c$ and $\bar{D}\Lambda_c$ 
  allows us to make predictions for each of the two different spin assignments, which turn out to be quite different, and then we make a call for the experimental determination of these widths that should settle clearly the spins of the two states. We also evaluate the mass splitting of the states due to the intermediate propagation of $\bar{D} \Sigma_c$ and $\bar{D}\Lambda_c$ states and find that the state with $1/2^-$ finds more attraction than the $3/2^-$, favoring again the $1/2^-$, $3/2^-$ assignment, but the amount of splitting found is small compared with experiment, pointing to extra elements responsible for the shift that merit being investigated. 

\section{Appendix A}
The integral of Eq.~(\ref{allt}) has two structures:
\begin{align}
   & \int\frac{d^3~q}{(2\pi)^3}~f(\Vec{q},\,\Vec{p}_1)q_i,\\\nonumber
   & \int\frac{d^3~q}{(2\pi)^3}~f(\Vec{q},\,\Vec{p}_1)q_iq_j.\\    
\end{align}
\par The first one is a vector and after integrating over $\Vec{q}$ can only be something {proportional to } $\Vec{p}_1$.
The second is a symmetric tensor {of rank two}, and after integrating over $\Vec{q}$ can give rise to two structures $\delta_{ij}$ and $p_{1i}~p_{1j}$. Thus, 
\begin{align}
   & \int\frac{d^3~q}{(2\pi)^3}~f(\Vec{q},\,\Vec{p}_1)q_i=a~ p_{1i}.   
\end{align}
\par Multiplying both sides by $p_{1i}$ and summing over $i$, we get 
\begin{align}
 a=  & \int\frac{d^3~q}{(2\pi)^3}~f(\Vec{q},\,\Vec{p}_1)\frac{\vec{q} \cdot \vec{p}_1}{\vec{p}_1^{~2}}.
\end{align}
\par Similarly
\begin{align}
   & \int\frac{d^3~q}{(2\pi)^3}~f(\Vec{q},\,\Vec{p}_1)q_iq_j=b~ {\vec{p}_1^{~2}}\delta_{ij}+c~p_{1i}~p_{1j},
\end{align}
multiplying by  $\delta_{ij}$  and summing over indices, and multiplying by  $p_{1i}~p_{1j}$  and summing over indices, we obtain two equations from where we deduce
\begin{align}
&  b=\int \frac{d^3 q}{(2 \pi)^3} \frac{1}{2 \vec{p}_1^{~4}}\left[\vec{q}^{~2} \vec{p}_1^{~2}-\left(\vec{q} \cdot \vec{p}_1\right)^2\right]~f(\Vec{q},\,\Vec{p}_1) , \\\nonumber
& c=\int \frac{d^3 q}{(2 \pi)^3} \frac{1}{2 \vec{p}_1^{~4}}\left[3\left(\vec{q} \cdot \vec{p}_1\right)^2-\vec{q}^{~2} \vec{p}_1^{~2}\right] ~f(\Vec{q},\,\Vec{p}_1).
\end{align}

\section{Acknowledgments}
This work is partly supported by the National Natural Science
Foundation of China under Grants  No. 12405089 and No. 12247108 and
the China Postdoctoral Science Foundation under Grant
No. 2022M720360 and No. 2022M720359. ZY.Yang and J. Song wish to thank support from the China Scholarship Council. This work is also supported by
the Spanish Ministerio de Economia y Competitividad (MINECO) and European FEDER funds under
Contracts No. FIS2017-84038-C2-1-P B, PID2020-
112777GB-I00, and by Generalitat Valenciana under con-
tract PROMETEO/2020/023. This project has received
funding from the European Union Horizon 2020 research
and innovation programme under the program H2020-
INFRAIA-2018-1, grant agreement No. 824093 of the
STRONG-2020 project. This work is supported by the Spanish Ministerio de Ciencia e Innovaci\'on (MICINN) under contracts PID2020-112777GB-I00, PID2023-147458NB-C21 and CEX2023-001292-S; by Generalitat Valenciana under contracts PROMETEO/2020/023 and  CIPROM/2023/59. 
This work is partly supported by the National Natural Science Foundation of China (NSFC) under Grants No. 12365019 and No. 11975083, and by the Central Government Guidance Funds for Local Scientific and Technological Development, China (No. Guike ZY22096024), the Natural Science Foundation of Guangxi province under Grant No. 2023JJA110076.

%\newpage
\bibliography{refs.bib} 
\end{document}